\documentstyle[11pt,newpasp,twoside,epsf]{article}

\begin{document}

\title{Dust Absorption and Emission in Galaxies at High and Low Redshifts}
\author{S. Michael Fall}

\affil{Space Telescope Science Institute, 3700 San Martin Drive, Baltimore,
MD 21218, USA}

\begin{abstract}
This article reviews three related topics: the extragalactic background 
light and its sources, evolution models for the dust absorption and 
emission in galaxies, and empirical constraints on these transfer 
processes in nearby starburst galaxies. 
It is intended that the material presented here will serve as an 
introduction to this Joint Discussion.
\end{abstract}

\section{Extragalactic Background Light}

A good place to begin our discussion of the effects of dust on the 
radiation from galaxies is with the extragalactic background light
(EBL). 
We denote the intensity of this isotropic radiation per unit frequency 
$\nu$ by $J_{\nu}$ and the integral of this over all $\nu$, the bolometric
intensity, by $J_{\rm bol}$.
By definition, $J_{\nu}$ excludes foreground radiation from the Milky 
Way and other nearby galaxies and the cosmic microwave background
radiation from the Big Bang, but it includes the radiation from 
everything in between.
A few years ago, we had only rough estimates and upper or lower limits
on $J_{\nu}$ at most wavelengths, but this situation has improved 
dramatically, thanks largely to {\it HST} and {\it COBE}.
We now know $J_{\nu}$ to an accuracy of a factor of two or better over
four decades in wavelength, from about 0.2 to 2000~$\mu$m.
This is shown in Figure~1 below. 

\begin{figure}
\plotfiddle{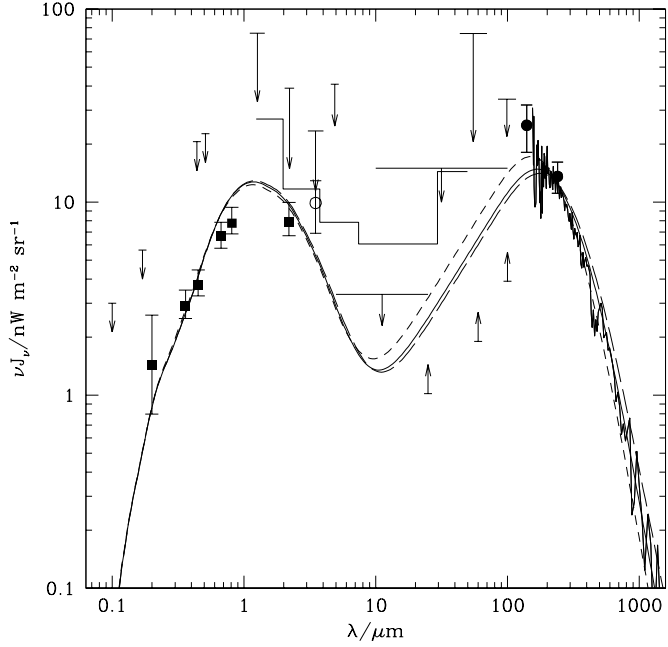}{3.2in}{0}{48}{48}{-155}{-95}
\caption{Spectrum of the extragalactic background light, $\nu J_{\nu}$
versus $\lambda$.
See Pei, Fall, \& Hauser (1999) for references to the observations.
The curves are from the global evolution models in the same paper.}
\end{figure}

Two interesting results follow directly from Figure~1.
First, emission by dust is important.
There is at least as much energy in the long-wavelength hump of the 
EBL spectrum ($\lambda \ga 10~\mu$m) as in the short-wavelength hump 
($\lambda \la 10~\mu$m).
Second, integrations under the smooth curves in Figure~1 give
\begin{equation}
J_{\rm bol}\approx50 \,\,{\rm nWm^{-2}sr^{-1}}.
\end{equation}
This is a summary record of the bolometric emissivity $E_{\rm bol}$ 
(power radiated per unit comoving volume) integrated over the age of 
Universe $t_0$:
\begin{equation}
J_{\rm bol} = {c\over4\pi}\int^{t_0}_0 {E_{\rm bol}\over{1+z}} dt.
\end{equation}
Equations~(1) and~(2) provide valuable constraints on possible sources 
of the EBL, independent of the complex radiative transfer within 
galaxies.

The most promising sources of the EBL are stars (nuclear energy) and
AGN (accretion onto black holes, gravitational energy).
Equation~(2) enables us to express these contributions to the EBL in 
terms of the present comoving densities of stars and black holes
(normalized by the critical density).
Recent estimates give $\Omega_{\rm S}\approx4\times10^{-3}$ and
$\Omega_{\rm BH}\approx2\times10^{-5}$ for $H_0 = 
70$~kms$^{-1}$Mpc$^{-1}$.
For stars, we obtain from population synthesis models
\begin{equation}
J_{\rm S} \approx 50 f (\Omega_{\rm S}/4\times10^{-3}) \,\,
{\rm nWm^{-2}sr^{-1}},
\end{equation}
where $f\approx1$ depends on the history of star formation and the 
initial mass function.
For AGN, we obtain
\begin{equation}
J_{\rm AGN} \approx 6 (\epsilon/0.05) [3/(1+z_{\rm acc})] 
(\Omega_{\rm BH}/2\times10^{-5}) \,\,{\rm nWm^{-2}sr^{-1}},
\end{equation}
where $\epsilon$ is the efficiency of energy conversion (gravitational
to radiative) and $z_{\rm acc}$ is the effective redshift of accretion.
According to these estimates, which have substantial uncertainties, stars 
alone could produce the entire EBL, while AGN may make a significant but
smaller contribution.
Similar conclusions have been reached by Fabian \& Iwasawa (1999) and
Madau \& Pozzetti (2000).

\section{Evolution Models}

Several methods have been developed to interpret or predict the 
appearance of galaxies at different redshifts and their contribution
to the cosmic emissivity and EBL.
In many cases, the absorption and reradiation of starlight by dust
is governed by an effective optical depth $\tau_{\rm dust}$ that is 
related to the metallicity, column density, and mass of the interstellar
medium, $Z_{\rm ISM}$, $N_{\rm ISM}$, $M_{\rm ISM}$, and the size of 
a galaxy $R_{\rm gal}$ by expressions of the form
\begin{equation}
\tau_{\rm dust} \propto Z_{\rm ISM} N_{\rm ISM} 
\propto Z_{\rm ISM} (M_{\rm ISM}/R_{\rm gal}^2).
\end{equation}
The mass and metallicity of the ISM are then related to the prior
history of star formation by the equations of galactic chemical 
evolution, with or without inflow or outflow terms.
Most recent models can be classified into three types: ``backward'' 
evolution models, ``forward'' evolution models, and ``global'' 
evolution models. 

In backward evolution models, the present-day population of galaxies
is evolved into the past using plausible rules for the histories of
star formation, gas consumption, and so forth in galaxies of different
types (see Franceschini et al. 1994 and references therein).
Interactions and merging of galaxies are usually ignored in this 
approach.
In forward evolution models, initial distributions of dark matter
and baryons are evolved into the future using plausible rules for the 
merging of halos, inflow and outflow of gas, and so forth (see Granato 
et al. 2000 and references therein).
These effects are usually treated by semi-analytical techniques in the 
framework of the cold dark matter cosmogony.
Since this approach includes many different processes, it requires a 
large number of assumptions and parameters.

Global evolution models are based on the combined or average properties 
of galaxies (see Pei, Fall, \& Hauser 1999 and references therein).
The quantities of interest here are the mean comoving densities of 
stars, ISM, metals, and dust ($\Omega_{\rm S}$, $\Omega_{\rm ISM}$, 
$\Omega_{\rm M}$, and $\Omega_{\rm D}$) and the mean comoving 
emissivities of stars and dust ($E_{{\rm S}\nu}$ and $E_{{\rm D}\nu}$).
These quantities are governed by simple, conservation-type equations, 
analogous to the equations of galactic chemical evolution.
The input to and output from the global evolution models include 
both emission and absorption-line observations, i.e., information
about both the stellar and interstellar contents of galaxies.
The advantages of this approach are that it is based directly on global 
quantities, requires relatively few assumptions and parameters, and 
relates the emission histories of galaxies to their absorption histories.
A limitation of the global evolution models is that they do not describe
the properties of individual galaxies and hence cannot be compared with
luminosity functions and number counts.
Moreover, the accuracy that is currently achievable is limited by 
small-number statistics in the existing samples of damped Ly$\alpha$ 
absorbers.

\section{Nearby Starburst Galaxies}

To determine the rate of stellar nucleosynthesis or black hole accretion
in a galaxy, one must measure its bolometric luminosity.
However, this is not currently practicable for most of the apparently 
faint galaxies at high redshifts. 
Most observations of high-redshift galaxies are restricted to rest-frame
UV and optical wavelengths, which miss the light that is absorbed and 
reradiated by dust.
Recent observations at 850~$\mu$m with SCUBA help, but much of the dust
emission is almost certainly at shorter wavelengths.
This situation has prompted several schemes to correct the observed
UV and optical fluxes to bolometric fluxes.
One of the most promising of these was discovered by Meurer et al. 
(1995) and refined by Meurer, Heckman, \& Calzetti (1999).

The Meurer et al. relation is based on the observed properties of 
UV-selected, nearby starburst galaxies. 
These are of special interest because they may be low-redshift analogs 
of the high-redshift ($z\approx3$--4) galaxies revealed by the Lyman-break 
technique (Steidel et al. 1996).
In the nearby sample, Meurer et al. found a relatively tight correlation 
between the ratio of FIR and UV fluxes, $F_{\rm dust}/F_{1600}$, and the 
UV spectral slope, $\beta$.
The correlation may also hold for the Lyman-break galaxies, although
the observational tests for this are not strong (Adelberger \& Steidel 
2000).
The Meurer et al. relation is useful because, for galaxies that obey 
it, one can infer the FIR flux $F_{\rm dust}$ and hence the bolometric 
flux from observations in the rest-frame UV ($F_{1600}$ and $\beta$).
In this way, Meurer el al. (1999) and Adelberger \& Steidel (2000) 
have estimated bolometric correction factors of about 5 between the 
UV-only and total star formation rates in Lyman-break galaxies at 
$z\approx3$--4.

The existence of the Meurer et al. relation is closely linked to
the existence of a ``universal'' effective absorption curve, although
not quite the same as the one proposed by Calzetti, Kinney, \&
Storchi-Bergmann (1994).
In this case, the parameter along the sequence is the effective 
optical depth at some fiducial wavelength and hence the overall dust 
content of the galaxies.
Another curious fact about the UV-selected, nearby starburst galaxies
is that the apparent absorption inferred from the H$\alpha$/H$\beta$
ratio is typically twice that inferred from the UV spectral slope
$\beta$ (Calzetti et al. 1994).
In an effort to understand these observations, Charlot \& Fall (2000)
have constructed some simple models in which stars are born in dusty
clouds with finite lifetimes.
As a result, young, UV-bright stars are more obscured than old,
UV-faint stars.
In particular, the Meurer et al. relation, and the other average
spectral properties of the UV-selected, nearby starburst galaxies, 
can be reproduced if the effective absorption is proportional to 
$\lambda^{-0.7}$ and is three times higher for stars younger than 
$10^7$~yr than it is for older stars.

\end{document}